\documentclass[12pt,letterpaper]{iopart}

\usepackage{graphicx}
\usepackage{dcolumn}

\usepackage[sort&compress,numbers]{natbib}

\newcommand{\ie}{\textit{i.e.}}
\newcommand{\eg}{\textit{e.g.}}

\newcommand{\boldvec}[1]{\ensuremath{\mathbf{\vec{#1}}}}
\newcommand{\gae}{%
  \ensuremath{\lower 2pt \hbox{%
    $\, \buildrel {\scriptstyle >}\over {\scriptstyle \sim}\,$}%
    }
  }
\newcommand{\lae}{%
  \ensuremath{\lower 2pt \hbox{%
    $\, \buildrel {\scriptstyle <}\over {\scriptstyle \sim}\,$}%
    }
  }

\newcommand{\TA}{\ensuremath{\mathrm{TA}}}   

\newcommand{\refeqn}[2][eqn:]{(equation~\ref{#1#2})}
\newcommand{\refEqn}[2][eqn:]{Equation~(\ref{#1#2})}
\newcommand{\reftab}[2][tab:]{table~\ref{#1#2}}

\newcommand{\reffig}[2][fig:]{figure~\ref{#1#2}}
\newcommand{\refFig}[2][fig:]{Figure~\ref{#1#2}}
\newcommand{\refsec}[2][sec:]{section~\ref{#1#2}} 

\begin{document}




\title[Optical depth to the LMC for a triaxial Milky Way Spheroid]{%
       New models for a triaxial Milky Way Spheroid and effect on the
       microlensing optical depth to the Large Magellanic Cloud}

\author{%
    C. Savage$^1$,
    H.J. Newberg$^2$,
    K. Freese$^1$
    and P. Gondolo$^3$}

\address{$^1$
    Michigan Center for Theoretical Physics,
    Department of Physics,
    University of Michigan,
    Ann Arbor, MI 48109
}
\address{$^2$
    Department of Physics,
    Applied Physics and Astronomy,
    Rensselaer Polytechnic Institute,
    Troy, NY 12180
}
\address{$^3$
    Physics Department,
    University of Utah,
    Salt Lake City, UT 84112
}
\eads{
    \mailto{cmsavage@umich.edu},
    \mailto{heidi@rpi.edu},
    \mailto{ktfreese@umich.edu}
    and \mailto{paolo@physics.utah.edu}
}

\date{\today}

\begin{abstract}

  We obtain models for a triaxial Milky Way spheroid based on data
  by Newberg and Yanny.  The best fits to the data occur for a
  spheroid center that is shifted by 3kpc from the Galactic Center.
  We investigate effects of the triaxiality on the microlensing
  optical depth to the Large Magellanic Cloud (LMC).  The optical
  depth can be used to ascertain the number of Massive Compact Halo
  Objects (MACHOs); a larger spheroid contribution would imply fewer
  Halo MACHOs.  On the one hand, the triaxiality gives rise to more
  spheroid mass along the line of sight between us and the LMC and
  thus a larger optical depth.  However, shifting the spheroid center
  leads to an effect that goes in the other direction: the best fit to
  the spheroid center is {\it away} from the line of sight to the LMC.
  As a consequence, these two effects tend to cancel so that the
  change in optical depth due to the Newberg/Yanny triaxial halo is at
  most 50\%.  After subtracting the spheroid contribution in the four
  models we consider, the MACHO contribution (central value) to the
  mass of the Galactic Halo varies from $\sim (8-20)\%$ if all excess
  lensing events observed by the MACHO collaboration are assumed to be
  due to MACHOs.  Here the maximum is due to the original MACHO
  collaboration results and the minimum is consistent with 0\% at the
  1$\sigma$ error level in the data.

\end{abstract}

\submitto{Journal of Cosmology and Astroparticle Physics}

\maketitle


\section{\label{sec:Intro} Introduction}

The mass of galaxies, including our Milky Way, is composed primarily
of unknown dark matter.  While the dark matter may consist primarily
of elementary particles (such as axions or Weakly Interacting Massive
Particles), current data indicate that up to 20\% of it may be
composed of Massive Compact Halo Objects (MACHOs).  From a combination
of theoretical arguments and data, low mass stars and brown dwarfs
have been shown to make up no more than a few percent of the Halo
\citep{Graff:1996rz,Graff:1995ru,Fields:1999ar,Freese:2000vi,
Alcock:2000ph,Lasserre:2000xw,Afonso:2002xq}.
Viable MACHO candidates include primordial black holes or white dwarfs.
[Constraints on  white dwarfs as explanations for MACHO events
 have previously been discussed in \citet{Fields:1998ja,Graff:1999gv,
 Fields:1999ar}.]

As noted by \citet{Paczynski:1985jf}, these MACHOs may be detected
through gravitational microlensing.  A survey of a large number of stars
for microlensing events can indicate an excess optical depth due to the
MACHOs.  Experiments such as
MACHO \citep{Alcock:2000ph},
EROS \citep{Lasserre:2000xw,Afonso:2002xq},
OGLE \citep{Udalski:1994ei,Udalski:2000nn},
AGAPE \citep{Ansari:1996xh,CalchiNovati:2001jk},
MOA \citep{Sumi:2002ez,Sumi:2002wg},
MEGA \citep{deJong:2003wc},
and \citet{Baltz:2003ds}
have carried out searches for MACHOs by searching for microlensing
events in large populations of stars such as the Large Magellanic Cloud
(LMC), Small Magellanic Cloud (SMC), Galactic bulge, M31, and M87.
Lensing events have been detected toward the LMC in greater numbers
than would be expected for the assumed stellar populations, indicating
the possible presence of MACHOs.  These experiments have found that as
much as 20\% of the Halo of the Milky Way may be composed of a MACHO
component.

However, the interpretation of these microlensing results as a new
MACHO component requires one to subtract off the contribution from
known stellar populations.  The stellar contribution in turn relies
upon the details of the galactic structure-- an incorrect model
for the stellar distribution may yield an underestimate for this
contribution and, hence, lead to some lensing events being mistakenly
attributed to MACHOs.  One galactic component contributing to the LMC
optical depth is the spheroid, or stellar halo, a (roughly spherical)
distribution of older stars with a density that falls off more rapidly
than the dark halo.  The traditional assumption, based upon star counts,
is that this spheroid is light and has either a spherical or axial
(flattened) shape; the traditional spheroid contributes little to the
LMC optical depth.  A heavy spheroid has also been considered in the
past \citep{Giudice:1993bp}.

Prior to this work, there have been two quite divergent results
for the optical depth due to the spheroid. The MACHO experiment
used a spheroid optical depth that is 1/40 of the total optical
depth to the LMC.  Another model, the OC1 result discussed below,
found a spheroid optical depth that is 1/3 of the total to the LMC.

In this paper we will investigate the consequences of a triaxial
shape for the Milky Way spheroid.  \citet{Newberg:2005ih,Newberg:2005mu}
have used $F/G$ turnoff star counts to find a triaxial spheroid.
In this paper, we will obtain models for the spheroid based on the
Newberg/Yanny data.  We will also examine how triaxiality affects the
optical depth and determine if triaxial models can account for the
excess optical depth toward the LMC.

The paper is organized as follows: in \refsec{OpticalDepth}, we will
review the determination of the optical depth to the LMC.  In
\refsec{Models}, we will present two existing models of the spheroid
as well as new triaxial models based upon $F/G$ turnoff star counts.
The importance of various factors (the triaxiality, position of the
spheroid, and shape and normalization of the density profile) on the
optical depth is discussed in \refsec{Factors}.  Results are
presented in \refsec{Results}, discussed in \refsec{Discussion}, and
summarized in \refsec{Summary}.

\section{\label{sec:OpticalDepth} Optical Depth}

In microlensing experiments, a telescope on Earth monitors millions of
stars in a nearby galaxy, \eg\ the LMC.  Occasionally one of these
stars will be lensed by an intervening mass, \eg\ a MACHO or a star in
the spheroid.  The distant source star will look brighter during the
time period that the intervening mass passes across the line of sight
between the telescope and the source star.  The optical depth is
roughly the probability of finding that the source star is lensed.
More precisely, in the low optical depth limit that applies to
observations of the Milky Way and other nearby galaxies, the optical
depth is the probability that a source object (\eg\ a star in the LMC)
will have its intensity amplified by more than a factor of 1.34 due to
gravitational microlensing by an intervening object.  It has been shown
(see the reviews of \citet{Paczynski:1996nh} and \citet{Roulet:1996ur})
that the optical depth $\tau$ is
\begin{equation} \label{eqn:od}
  \tau = \frac{4 \pi G}{c^2} D_s^2
         \int_0^{1} \mathrm{d}x \, x \, (1 - x) \,\,
           \rho\left[ (1 - x) \boldvec{r}_\odot + x \boldvec{r}_s
               \right] .
\end{equation}
Here $\boldvec{r}_s$ is the location of the source star,
$\boldvec{r}_\odot$ is the location of the Sun (we will approximate
our position as being at the Sun), and $\rho(\boldvec{r})$ is the mass
density of lensing objects at a position $\boldvec{r}$.  As density
profiles discussed here depend on the radial distance from the center
of the spheroid (traditionally the same as the Galactic Center), we
will choose the origin of the coordinate system to lie at the center
of the spheroid.  The coordinate $x$ measures the fractional distance
along the line of sight from the Sun toward the source star, where
$D_s$ is the distance between us and the source object.

Typically lenses can be expected to traverse the line of sight to
source stars, resulting in temporary amplification.  The optical depth
toward the LMC can then be determined by examining a large number of
LMC stars; the average proportion of stars lensed at any one time is
the optical depth.  Lensing objects with masses $\sim (10^{-6}$ -
$10^2) \, M_\odot$ have event timescales on the order of days or
months and are observable by the current experiments.  The LMC,
located at $(d_{LMC}, \, b_{LMC}, \, \ell_{LMC}) = (50 \,
\mathrm{kpc}, \, -32.8^\circ, \, 281^\circ)$, provides a nearby source
of stars with a line of sight off the galactic plane that does not
pass through the Galactic Center (the advantage of this line of sight
is that it avoids the densest stellar regions which would overpower
the MACHO signal of interest).
Several stellar populations will contribute to the optical depth to LMC
source stars:  the Milky Way disk (both thin and thick components), the
Milky Way spheroid, and the LMC disk itself.  MACHOs in the Milky Way
and LMC haloes will also contribute to the optical depth.  

The MACHO collaboration has determined an optical depth to the LMC of
$\tau_2^{400} = 1.1_{-0.3}^{+0.4} \times 10^{-7}$ (criteria set A, 13
events) and $\tau_2^{400} = 1.3_{-0.3}^{+0.4} \times 10^{-7}$
(criteria set B, 17 events) \citep{Alcock:2000ph}; these values of the
optical depth are for events with durations between 2 and 400 days.  
Their prediction for ``known'' stellar populations (MW disk, MW
spheroid, LMC disk) is $\tau_{stars} = (0.24-0.36) \times 10^{-7}$
($0.020 \times 10^{-7}$ from the spheroid), only a fraction of the
observed value. The optical depth has a noticeable excess
of $\sim 1 \times 10^{-7}$ from the known contributions, with
three possibilities (or combination thereof) for this discrepancy: (1)
MACHOs, which can contribute up to 20\% of the dark matter halo; (2)
backgrounds, such as variable stars, misinterpreted as lensing events;
and (3) incorrect stellar population models, yielding ``known''
contributions smaller than the actual stellar contributions.  Our
examination here involves the third case: we will investigate whether
a triaxial spheroid will have a larger contribution than the spherical
models and account for part of this excess.

\section{\label{sec:Models} Spheroid Models}

Modeling the spheroid has presented some difficulty.  The spheroid
density is expected to fall off faster than the dark halo, with a
power law dependence far away from the Galactic Center roughly
$\propto r^{-n}$ with $n \sim 3$.  However, the inner galaxy profile
and total mass (alternatively, local density) are not as well known.
Star counts, including surveys of local high velocity stars, and
dynamical models based upon rotation curves lead to very different
spheroid masses.  In this section, we will present a model for each of
these cases.  In addition, we will examine new models due to
\citet{Newberg:2005ih,Newberg:2005mu}, based upon $F/G$ turnoff star
 counts.

\subsection{\label{sec:MACHO} Star Counts and High Velocity Stars:
            ``MACHO model''}

In principle, the profile of the spheroid could be determined if all
of its stars (and other composing objects) could be directly observed.
In practice, this becomes difficult as these objects may be too faint
to see and may be difficult to distinguish
from other stellar populations-- the local spheroid density is much
smaller ($\sim 0.1\%$) than that of the disk.  Models can be
constructed by observing a distinguishable population of spheroid
stars and extrapolating to a total density by using the Mass Function
(MF)~\citep{Scalo:1986ja,Gould:1996hv,Gould:1998aw,
Graff:1996rz,Graff:1995ru,Chabrier:2003ki}.
Spheroid stars can be distinguished due to their large
relative velocity (disk stars tend to have a small peculiar velocity as
they rotate about the center of the galaxy along with the
Sun) or their chemical composition; star counts at high galactic
latitudes (away from the disk) can also be used
\citep{
  Schmidt:1975,Bahcall:1986,Dahn:1995,Gizis:1998wh,Cooke:2000,Gould:2002hf,
  Norris:1985,Chiba:2000vu,
  Bahcall:1980fb,Gould:1998aw,Digby:2003tt,Lemon:2003pg,
  Bahcall:1983}.

One assumption made in these type of determinations is an extrapolation
of stellar mass functions to the unobservable low mass end of the
spectrum (\eg\ brown dwarfs); typical extrapolations predict only a
small portion of the mass density is composed of brown dwarfs.
Atypical models that propose larger amounts of brown dwarfs are ruled
out by a combination of theoretical arguments and data
\citep{Graff:1996rz,Graff:1995ru,Fields:1999ar,Freese:2000vi,
Alcock:2000ph,Lasserre:2000xw,Afonso:2002xq}.  In particular, the time
scales of observed MACHO events are inconsistent with large numbers of
very light or very heavy objects, requiring mainly objects of comparable
mass to main sequence stars.

The MACHO collaboration uses a $r^{-3.5}$ power law model based partly
upon the kinematically selected observations~\citep{Alcock:2000ph},
\begin{equation} \label{eqn:MACHOProfile}
  \rho(r) = \rho_{S0} \left( \frac{r}{R_0} \right)^{-3.5} ,
            \quad \quad \textrm{(MACHO)}
\end{equation}
with a local spheroid density (in the solar vicinity) of 
\begin{equation} \label{eqn:MACHOnorm}
  \rho_{S0} = 1.18 \times 10^{-4} M_{\odot} \, \mathrm{pc}^{-3} .
\end{equation}
Here $R_0$ is the distance between the Sun and the Galactic Center.  We
will refer to this power law fit as the MACHO profile.  Due to the
statistical and systematic errors and assumptions made in the
determination of this local density used by the MACHO Collaboration,
it is possible that it is off by at most a factor of 2
(W.~Sutherland 2005, private communication).

Other observations further constrain the local density.  The
kinematically selected observations lead to estimates of the local
spheroid density of $\rho_{S0} \approx (1-3) \times 10^{-4} M_{\odot}
\, \mathrm{pc}^{-3}$, where these values may contain significant
statistical and systematic errors.  The high galactic latitude results
of \citet{Gould:1998aw} yield a noticeably smaller estimate of $6.4
\times 10^{-5} M_{\odot} \, \mathrm{pc}^{-3}$; the latter analysis
relies upon an extrapolation from distant to nearby densities that
depends on the spheroid model and could be an underestimate.  A recent
paper by \citet{Gould:2002hf}, with a large number of kinematically
selected stars, presents a luminosity function from which a more precise
value of the local density might possibly be obtained.  However,
systematics in the data  and errors in extrapolating the luminosity
function to a local density must then be considered; this analysis
has not yet been done to our knowledge.  Even with all the uncertainties
in all these measurements taken into account, the values of the local
densities used by the MACHO group or obtained by these other
measurements are still in severe disagreement with the local density
used in the OC1 model described in the next section: OC1 gives a value
of the local density that is an order of magnitude higher than that used
by the MACHO group.  Later, we will also examine new work based upon
$F/G$ turnoff star counts \citep{Newberg:2005ih,Newberg:2005mu} where
the estimates of the local density are lower.

However, it may not be appropriate to compare the local density, as
measured by counting stars in the solar neighborhood, with the local
stellar density of stars calculated from the Newberg/Yanny triaxial
spheroid.  If the spheroid is lumpy or contains multiple components,
then the local density could be significantly higher than interpolated
from the triaxial spheroid, which was fit to a selection of halo stars
which avoided known halo substructures.  In \refsec{Hernquist}, we
calculate the contribution to the local density expected from each of
four best fit $F/G$ spheroid models.  In \refsec{DensityDisc}, we
further discuss the local density of all the spheroid models.

\subsection{\label{sec:OC} Galactic Dynamics: ``OC1 Model''}

Alternatively, models can be generated based upon dynamical
measurements, using simulations to match observed rotation
curves~\citep{Caldwell:1981rj,Ostriker:1982,Rohlfs:1988}.
Such models tend to lead to much heavier spheroids than those based
upon star counts, by as much as a factor of 10.
\citet{Giudice:1993bp}
examined several different models for the spheroid; 
here, we will use the one they refer
to as OC1 (proposed by \citet{Ostriker:1982}).
This model is based upon a variety of observational data, but depends
significantly upon rotation curves for $R < R_0$ as determined from
infrared surveys.  For the radii relevant to the LMC line of sight,
\begin{equation} \label{eqn:OC1Profile}
  \rho(r) = \rho_{S0} \left( \frac{r}{R_0} \right)^{-3} ,
\end{equation}
with a local spheroid density of $\rho_{S0} = 1.28 \times 10^{-3}
M_{\odot} \, \mathrm{pc}^{-3}$, an order of magnitude larger than that
used by MACHO.  We note that this ``local density'' is obtained
from the modeling, which is designed to match a number of
data sets, rather than directly from local observations.

\subsection{\label{sec:Hernquist} $F/G$ Turnoff Star Counts:
            ``Power Law Models'' and ``Hernquist Models''}

Recently, \citet{Newberg:2005ih,Newberg:2005mu} have provided new
models for the spheroid by examining $F/G$ turnoff stars using the Sloan
Digital Sky Survey (SDSS) \citep{York:2000gk,Abazajian:2004it}.
Models are constructed by taking the SDSS data, removing regions of
known clumps of stars and contamination (including streams and other
remnants), and assuming a smooth distribution for the remaining regions.
The number density of spheroid $F/G$ turnoff stars was fit to a modified
Hernquist profile~\citep{Hernquist:1990be}:
\begin{equation} \label{eqn:nHern}
  n_{F/G}(r) = \frac{C}{r^\alpha \, (a + r)^\delta} ,
\end{equation}
where $a$ is the core radius.
Newberg and Yanny considered (i) a true Hernquist profile, which has
$\alpha = 1$ and $\delta = 3$; and (ii) the pure power law case of
arbitrary $\alpha$ with $\delta = 0$.  We do note that more general
profiles were also considered, and the data actually fit a profile
very well for intermediate cases with a range of $\alpha$ and $\delta$
such that $\alpha + \delta \approx 4$ (this family of models is also
known as $\lambda$- \citep{Dehnen:1993} or
$\eta$-models \citep{Tremaine:1993qb}).  Table I shows the two cases
considered by Newberg and Yanny. Two fits with pure Hernquist are
shown in the table as well as two fits to a pure power law with
$\alpha \approx 3$, similar to the power law MACHO and OC1 profiles.
The Hernquist profile provides a better fit to the data, but the power
law is included for comparison.  In total there are 14 parameters to
fit, which we now discuss.  The parameter $M_f$, the peak of the
absolute magnitude distribution of the turn-off stars, is fixed to 4.2
in obtaining the other parameters.


\newlength{\daggerlength}
\settowidth{\daggerlength}{$^{\dagger}$}
\newcommand{\emphparam}[1]{\hspace{\daggerlength}$\,$\textbf{#1}$\,^{\dagger}$}

\begin{table*}
  \caption[Model Parameters]{
    Several fits of $F/G$ turnoff star counts to the modified Hernquist
    profile (\refEqn{nHern}).
    Parameters denoted by ($\dagger$) have been fixed; all other
    parameters are permitted to vary.
    Models H1 and H2 have true Hernquist profiles, while P1 and P2 have
    power law profiles.  H1 \& P1 have the spheroid center fixed to
    lie at the Galactic Center; the center is free to move in H2 \& P2.
    The fourteen fit parameters are described in the text.
    }
  \label{tab:TAParams}
  \begin{indented}
  \item[]
  \begin{tabular}{lccccc}
    \hline\hline
    Model & &
      H1 & H2 & P1 & P2 \\
    \hline
    Magnitude & $M_f$ &
      \emphparam{4.2} & \emphparam{4.2} &
      \emphparam{4.2} & \emphparam{4.2} \\
    Location & $R_0$ (kpc) &
      $8.5$ & \emphparam{8.0} & $8.5$ & \emphparam{8.0} \\
    & $\Delta x$ (kpc) &
      \emphparam{0} & $0.1$ & \emphparam{0} & $0.2$ \\
    & $\Delta y$ (kpc) &
      \emphparam{0} & $3.5$ & \emphparam{0} & $2.9$ \\
    & $\Delta z$ (kpc) &
      \emphparam{0} & $0.1$ & \emphparam{0} & $0.0$ \\
    Profile & $\alpha$ &
      \emphparam{1} & \emphparam{1} & $3.0$ & $2.9$ \\
    & $\delta$ &
      \emphparam{3} & \emphparam{3} & \emphparam{0} & \emphparam{0} \\
    & $a$ (kpc) &
      $14$ & $15$ & -- & -- \\
    & $C$ (kpc$^{(\alpha+\delta-3)}$) &
      $2.09 \times 10^8$ & $1.55 \times 10^8$ &
      $2.16 \times 10^6$ & $1.07 \times 10^6$ \\
    Triaxiality & $p$ &
      $0.73$ & $0.73$ & $0.72$ & $0.74$ \\
    & $q$ &
      $0.60$ & $0.67$ & $0.59$ & $0.66$ \\
    & $\theta$ &
      $70^\circ$ & $48^\circ$ & $72^\circ$ & $52^\circ$ \\
    & $\phi$ &
      $-4.5^\circ$ & $-8^\circ$ & $-4^\circ$ & $-6.5^\circ$ \\
    & $\xi$ &
      $14^\circ$ & $12^\circ$ & $14^\circ$ & $16^\circ$ \\
    Goodness of fit & \textit{d.o.f.} &
      13199 & 13197 & 13199 & 13197 \\
    & $\chi^2$/\textit{d.o.f.} &
      1.42 & 1.37 & 1.51 & 1.49 \\
    \hline
  \end{tabular}
  \end{indented}
\end{table*}

\subsubsection{\label{sec:Position} Position}

In both the true Hernquist and power law models, Newberg and Yanny have
examined two cases: (1) the spheroid center is fixed to coincide with
the Galactic Center but the distance $R_0$ from the Sun to the Galactic
Center is allowed to vary, and (2) $R_0$ is fixed but the spheroid
center is unconstrained and need not coincide with the Galactic Center.
The distance to the spheroid center, $R_0 + \Delta x$, is allowed to
vary in both cases.  For these two cases, the Hernquist models are
denoted by H1 and H2 and the power law models by P1 and P2, where index
1 refers to fixed spheroid center and 2 refers to unconstrained spheroid
center.  For the unconstrained case, we take $(\Delta x,\Delta y,\Delta z)$
to identify the position of the center of the spheroid relative to the
Galactic Center in galactocentric coordinates.  The best fit for the
center of the spheroid is found to lie several kpc from the Galactic
Center in the direction of the $y$ axis.

An important result here is the fact that this best fit position of the
center of the spheroid is \textit{away} from the line of sight to the
LMC, and hence would imply a lower spheroid contribution to the optical
depth.  In the next section we will see that triaxiality provides an
effect in the opposite direction.

\subsubsection{\label{sec:Triaxial} Triaxiality}

\begin{figure}
  \begin{indented}
  \item[] \includegraphics[keepaspectratio,width=1.00\columnwidth,
                           height=0.40\textheight]{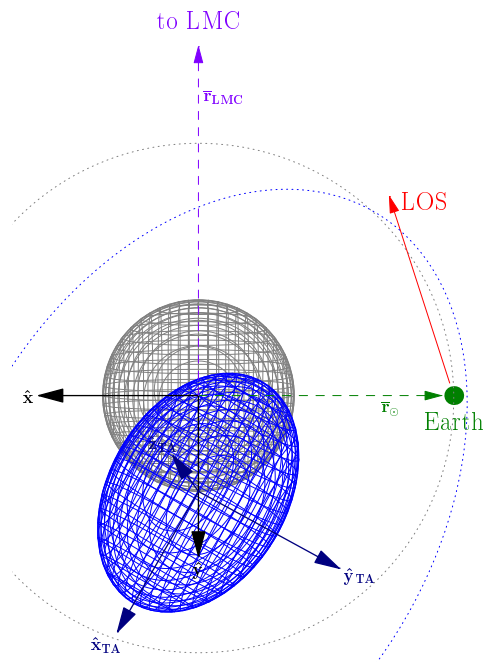}
  \end{indented}
  \caption[Spheroid Contours]{
    The location and shape of the spheroid
    found by Newberg and Yanny are shown in galactocentric coordinates
    (the $+z$ axis is out of the page).  The (grey) sphere represents
    the isodensity contour 3 kpc from the Galactic Center for a
    spherical density profile.  The (blue) ellipsoid indicates the
    equivalent triaxial isodensity contour (taking the spheroid mass to
    be the same as in the spherical case),
    with a shifted center; the triaxial axes are also shown.  The
    outer, dotted (gray) circle represents the extent of the spherical
    contour at 8 kpc; the dotted (blue) ellipse is the extent of the
    equivalent triaxial contour.  The line of sight (LOS) to the LMC
    near Earth passes farther within the triaxial contour than the
    spherical contour-- the densities along the LOS are therefore larger
    in this region for the triaxial case.
    }
  \label{fig:Contours}
\end{figure}

Previous observations and models have included the possibility of a
flattened (axial) spheroid.  The $F/G$ survey used by
Newberg and Yanny, however, is the first direct observation to strongly
indicate a triaxial distribution. [A triaxial inner halo has been
suggested by \citet{Larsen:1996} as a possible explanation for an
asymmetry they observed in their star counts.  A triaxial halo has also
been considered to explain indirect observations, such as rotation
curves; see, \eg, \citet{Blitz:1991}.]  The
triaxial isodensity contours fall on ellipsoids:
\begin{equation} \label{eqn:triaxialr}
  r_{\TA} = \sqrt{x^2 + \frac{y^2}{p^2} + \frac{z^2}{q^2}} ,
\end{equation}
with $p \approx 0.7$ and $q \approx 0.6$.  The primary axes are rotated
from the galactic coordinates by the angles $(\theta, \, \phi, \, \xi)
\approx (50$-$70^\circ, \, -(4$-$8^\circ), \, 12$-$16^\circ)$ -- the
major ($x$) and middle ($y$) axes lie near the plane of the galaxy, with
the major axis at approximately $60^\circ$ from our line of sight to the
Galactic Center [$(\theta,\phi,\xi)$ describe the orientation of the
spheroid axes and specify the following sequence of rotations: by
$\theta$ about $\mathbf{\hat{z}}$, then $\phi$ about
$\mathbf{\hat{y}'}$, and finally $\xi$ about $\mathbf{\hat{x}''}$].
\refFig{Contours} portrays the triaxial spheroid with a shifted center.

Note, with this definition, $r_{\TA}$ does not represent the
\textit{physical} distance to a given point, but is equal to the
distance along the major axis to the isodensity contour containing
that point (only for the specific case $p=q=1$ are these two distances
the same).  We will call this parameter $r_{\TA}$ the ``triaxial
radius''.  Then the local triaxial radius (the distance from the
spheroid center along the major axis to the isodensity contour
containing the Sun, i.e., the coordinate equivalent of $R_0$)
now becomes $R_{0,\TA} \approx 11.5$ kpc for the fixed position
models H1 \& P1 and $R_{0,\TA} \approx 9.5$ kpc for the shifted center
models H2 \& P2 (see \reftab{DerivedParams}).  The triaxial radius
$R_{0,\TA}$ is smaller in the cases with the shifted centers, even
though their physical distance is slightly larger, because the shift
places the local position nearer to the major axis.  In all cases, the
major axis passes somewhat between our local position and the LMC,
nearer to the local position. Hence the line of sight to the LMC can
be expected to pass through denser portions of the spheroid than in
the purely spherical case.

We emphasize here the important effect that the shape of the triaxial
spheroid tends to move matter into the line of sight between us and the
LMC (as compared to a spherical spheroid), so that the spheroid optical
depth would be higher and one would require fewer MACHOs.  In
\refsec{Results} we will see that the two competing effects of (i)
position of the center of the spheroid and (ii) triaxiality of the
spheroid tend to cancel.

Henceforth, in order to allow for triaxiality, we replace $r$ with $r_{\TA}$
as defined in Eq.(\ref{eqn:triaxialr}) in the density profiles.

\begin{table*}
  \caption[Derived Parameters]{
    Several derived parameters for the various models.  The local
    spheroid density is given using both the globular cluster (GC)
    and mass function (MF) estimates (described in the text).
    For comparison, the local density used by the MACHO Collaboration
    is $1.18 \times 10^{-4} M_{\odot} \, \mathrm{pc}^{-3}$.
    }
  \label{tab:DerivedParams}
  \begin{indented}
  \item[]
  \begin{tabular}{lccccc}
    \hline\hline
    Model & &
      H1 & H2 & P1 & P2 \\
    \hline
    Local Coordinate Distance  & $R_{0,\TA}$ (kpc) &
      $11.4$ & $9.5$  & $11.7$ & $9.7$  \\
    Local $F/G$ Number Density &
      $n_{F/G,0}$ ($\times 10^{-6} \mathrm{pc}^{-3}$) &
      $1.12$ & $1.11$ & $1.35$ & $1.47$ \\
    Local Density (GC) &
      $\rho_{S0}$ ($\times 10^{-5} M_{\odot} \, \mathrm{pc}^{-3}$) &
      $1.1$  & $1.1$  & $1.4$  & $1.5$  \\
    Local Density (MF)  &
      $\rho_{S0}$ ($\times 10^{-5} M_{\odot} \, \mathrm{pc}^{-3}$) &
      $4.5$  & $4.4$  & $5.4$  & $5.9$  \\
    \hline
  \end{tabular}
  \end{indented}
\end{table*}

\subsubsection{\label{sec:HernNorm} Density Normalization}

Assuming that the $F/G$ stars track the mass profile of the spheroid,
and using \refeqn{nHern} with the substitution $r \rightarrow 
r_{\TA}$ (as discussed in the previous section),
we can write the density profile of the spheroid as
\begin{equation} \label{eqn:HernProfile}
  \rho(r_{\TA}) = \rho_{S0} \left( \frac{R_{0,\TA}}{r_{\TA}} \right)^\alpha
                 \left( \frac{a + R_{0,\TA}}{a + r_{\TA}} \right)^\delta ,
\end{equation}
where the constant $C$ appropriate to the $F/G$ star density
has been replaced by appropriate terms with the local
triaxial radius.  We still need to determine the overall
normalization for the mass density profile for this Newberg/Yanny
profile.

The $F/G$ stars in the turnoff phase have masses $\sim 0.85
M_{\odot}$; these stars alone then contribute $\sim 1 \times 10^{-6}
M_{\odot} \, \mathrm{pc}^{-3}$ to the local density.  However, main
sequence stars, relics, brown dwarfs and other objects also contribute
to the spheroid mass and must be included.  We will estimate the total
local spheroid density $\rho_{S0}$ from the number density profile
normalization, $C$ of \refeqn{nHern}, using two different methods.

The first density determination is made by assuming the spheroid
population is similar to that of a globular cluster.  The Pal 5 globular
cluster, with a mass of $4.5$-$6 \times 10^3 M_{\odot}$
\citep{Odenkirchen:2002hk}, contains 620 stars with the same color cuts
as the spheroid data set (excluding the $r-i$ cut, which is not
significant in this case).  Off-field observations predict $\sim$32
background stars in the Pal 5 field, giving $\sim$588 stars above
background.  This yields a total stellar mass of approximately
$10 M_{\odot}$ per $F/G$ turnoff star.  Assuming the spheroid contains a
similar stellar population (an assumption which would need to be
justified), the spheroid would then have a total local density
$\rho_{S0} \sim 1 \times 10^{-5} M_{\odot} \, \mathrm{pc}^{-3}$; values
for the specific models are given in \reftab{DerivedParams}.

Alternatively, we can use the spheroid mass function \citep{Scalo:1986ja,
Gould:1996hv,Gould:1998aw,Graff:1996rz,Graff:1995ru,Chabrier:2003ki},
\begin{equation} \label{eqn:PDMF}  
  \frac{\mathrm{d}n}{\mathrm{d}M}(M) \propto
  \left\{
    \begin{array}{cl}
      M^{-1.1} , \quad & M \lae\ 0.71 M_\odot \\   
      M^{-2.7} ,  \quad & M \gae\ 0.71 M_\odot
    \end{array}
  \right.
\end{equation}
where this particular MF is taken from \citet{Gould:1998aw} (using the
binary corrected form); the general procedure here follows the one of
\citeauthor{Gould:1998aw}, albeit with an independent normalization.
The
$M^{-1.1}$ power law is thought to be valid for $0.09 M_\odot < M < 0.71
M_\odot$ and the $M^{-2.7}$ power law for $M \gae M_\odot$. We have
extended the low mass power law to all masses below $0.71 M_\odot$.  The
results are not particularly sensitive to where the power law changes in
the range $0.71$-$1 M_\odot$; here, we follow \citeauthor{Gould:1998aw}
and take the change to occur at $0.71 M_\odot$.  To determine the
normalization of the MF, we assume the spheroid stars were predominantly
formed at the same time, during the formation of the spheroid
$\sim 10^{10}$ years ago; then the stars currently in the turnoff phase
(lasting $\sim 3 \times 10^{9}$ years) correspond to a range of masses
with width $\Delta \tilde M \approx 0.07 M_\odot$ centered at
$\tilde M \approx 0.85 M_\odot$.  Though not all star formation occurred
at the same time, the number of $F/G$ turnoff stars should still give a
reasonable estimate for the number of stars in this range, so that
\begin{equation} \label{eqn:PDMFRange}
  n_{F/G} \approx \frac{1}{2}
                  \frac{\mathrm{d}n}{\mathrm{d}M}(\tilde M)
                  \, \Delta \tilde M ,
\end{equation}
where the extra factor of $1/2$ is due to the fact that the selection of
$F/G$ stars in the data corresponds to approximately the latter half of
the turnoff phase (color cuts used to obtain this selection are meant to
avoid contamination of the sample from other stellar populations).
Using \refeqn{PDMFRange} to fix the normalization in \refeqn{PDMF} and
noting that progenitors with $M \gae 0.9 M_\odot$ have mainly become
white dwarfs with $M \approx 0.6 M_\odot$, we integrate \refeqn{PDMF},
\begin{equation} \label{eqn:PDMFIntegral}
  \rho_{S0} = \int \mathrm{d}M \, M \, \frac{\mathrm{d}n}{\mathrm{d}M} ,
\end{equation}
to find the spheroid contains about $40 M_\odot$ of mass per $F/G$
turnoff star, or $\rho_{S0} \sim 5 \times 10^{-5} M_{\odot} \,
\mathrm{pc}^{-3}$.  This value varies by $\lae$ 20\% if the MF power law
change in \refeqn{PDMF} is taken to occur in the range [0.71,1.0]
$M_\odot$.  Values for the specific models are given in
\reftab{DerivedParams}.

The MF estimate of the local density is $\approx$4 times larger than the
globular cluster estimate, but both are below that determined by most
other models, although these estimates are not as rigorous and should be
regarded as rough estimates.  The exception is the high galactic
latitude measurements of \citet{Gould:1998aw}, which provides
a very similar estimate of the local density ($6.4 \times 10^{-5}
M_{\odot} \, \mathrm{pc}^{-3}$) as our MF estimate (while we use the
same MF as \citeauthor{Gould:1998aw}, we remind the reader that we have
independently estimated the normalization of this MF).  But, as our
estimates are non-rigorous, this similarity may only be a coincidence.
A discussion of the local density can be found in \refsec{Factors}.

We will henceforth use three possible values for the local
density:
\begin{itemize}
  \setlength{\itemsep}{0pt}
  \item[(i)]   MACHO: $\rho_{S0} = 1.18 \times 10^{-4} M_\odot \,
               \mathrm{pc}^{-3}$,
  \item[(ii)]  GC: $\rho_{S0} \sim 1 \times 10^{-5} M_\odot \,
               \mathrm{pc}^{-3}$, with more precise values given in
               \reftab{DerivedParams},
  \item[(iii)] MF: $\rho_{S0} \sim 5 \times 10^{-5} M_\odot \,
               \mathrm{pc}^{-3}$, with more precise values given in
               \reftab{DerivedParams}.
\end{itemize}

\section{\label{sec:Factors} Importance of Various Factors for Optical
         Depth}

The optical depth results depend on several factors: the triaxiality,
the position, the density profile, and the normalization (\ie\ total
mass or local density) of the spheroid.  We discuss each of these.

\subsection{\label{sec:TriaxialDisc} Triaxiality}

\begin{figure*}
  \begin{indented}
  \item[] \includegraphics[width=0.80\textwidth]{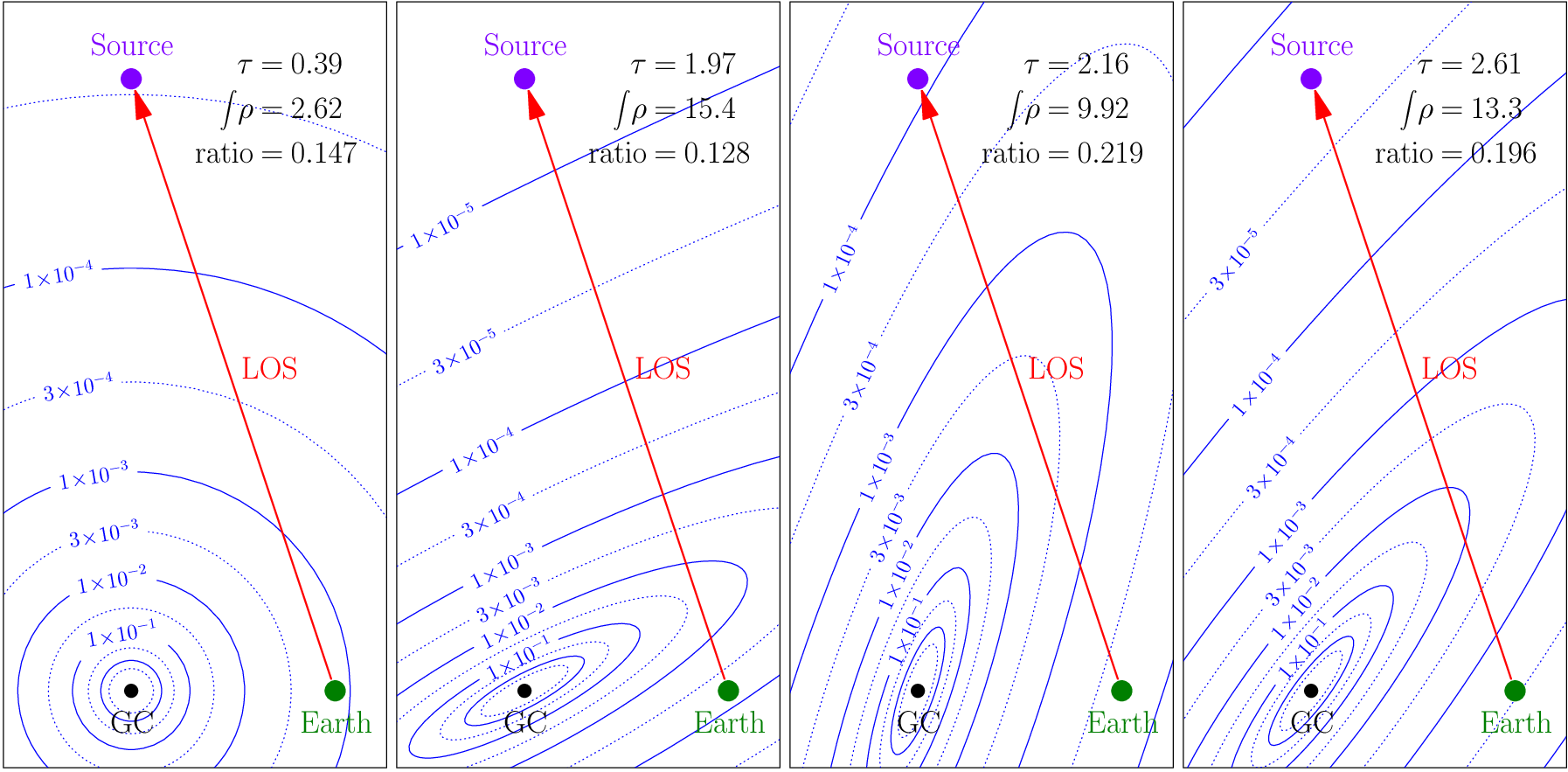} \\
  \hspace*{0.1\textwidth}\makebox[0pt][c]{(a)}
  \hspace*{0.2\textwidth}\makebox[0pt][c]{(b)}
  \hspace*{0.2\textwidth}\makebox[0pt][c]{(c)}
  \hspace*{0.2\textwidth}\makebox[0pt][c]{(d)}
  \end{indented}
  \caption[Major Axis Direction]{
    The spheroid density contours and optical depth toward a source in
    the Milky Way for various spheroid configurations.  A nearer source
    than the LMC and an exaggerated triaxiality ($p\!=\!q\!=\!0.25$)
    have been chosen for clearer visualization; the major axis lies
    somewhere along the line of sight (LOS). We use the OC1 model with a
    fixed mass normalization as an illustration.  In the upper right
    hand corner of each panel, we list the optical depth $\tau$ (in
    units of $10^{-7}$), the integral of the density along the LOS
    ($\int \rho$, corresponding to the total intervening amount of
    matter, in arbitrary units), and the ratio of these two quantities
    (representing the lensing efficiency).  Panel~(a) shows the
    spherical ($p\!=\!q\!=\!1$) case, while Panels (b), (c), and~(d)
    show major axis directions that maximize the mass along the LOS,
    lensing efficiency, and optical depth, respectively.
    }
  \label{fig:MADirection}
\end{figure*}

A triaxial spheroid can lead to a larger optical depth to the LMC for
two reasons: (1) a larger amount of matter along the line of sight and
(2) more effective placement of the matter.  The line of
sight between us and the LMC does not pass much closer to the Galactic
Center than our current position.  Hence, for a spherical density
distribution, the spheroid density along the line of sight is largest
near us and falls off essentially the entire way to the LMC.  The
models discussed in this paper, however, indicate that the spheroid is
triaxial, with the LMC line of sight passing somewhere near the major
axis.  This means that the line of sight passes through contours of
higher density and the total matter along the line of sight is larger
than in the spherical case, leading to a larger optical depth.

Not only is the total amount of matter along the line of sight
important, but its location along that line is also significant.  
Lensing objects near the source or observer do not contribute as
significantly to the optical depth as lensing objects halfway in
between, as is apparent from the $x (1-x)$ factor in the integrand of
\refeqn{od}. Objects near the observer have small $x$ while objects
near the source have small $(1-x)$; the largest contribution to the
integral is for $x \sim 1/2$.  This factor has geometrical origins: a
lensing object near the source (or observer) must bend the light
through a greater angle than one appoximately halfway between.  Thus a
lensing object near the source (or observer) must be much nearer to
the line of sight to produce a microlensing event; such an object near
the source or observer has a smaller lensing cross-section.  Lensing
objects approximately halfway along the line of sight, where $x (1-x)$
is maximized, have the largest lensing cross-section and contribute
more to the optical depth.  For a spherical density distribution, the
spheroid density is largest near us ($x \approx 0$), where the
contribution to the optical depth is suppressed due to this
geometrical effect.  In the triaxial models discussed, the largest
densities are not at our location, but farther away where spheroid
objects are more likely to produce lensing events.  Even if the total
amount of matter along the line of sight were the same, the triaxial
models in this paper would yield a higher optical depth than spherical
ones.

We demonstrate these effects with an example shown in
\reffig{MADirection}.  For clarity, we examine the optical depth
toward a source in the Milky Way nearer than the LMC and take an
exaggerated triaxiality ($p\!=\!q\!=\!0.25$) in the OC1 model (using a
fixed mass normalization).  In this figure, we show the spheroid
density contours and optical depth toward the source (in units of
$10^{-7}$) for various spheroid configurations where the major axis is
fixed to lie somewhere along the line of sight (LOS).  In addition to
the optical depth, each panel shows the integral of the density along
the LOS ($\int \rho$, corresponding to the total amount of intervening
matter, in arbitrary units) and the ratio of these two quantities
(representing the lensing efficiency).  Panel~(a) shows the spherical
($p\!=\!q\!=\!1$) case.  Panel~(b) shows a triaxial spheroid with the
major axis direction chosen to maximize the amount of matter along the
LOS.  Note the LOS passes through much higher density contours that
the spherical case, with $\sim$6 times the mass along the LOS.  Most
of this additional mass, however, is near to Earth, where $x$ is
small, so is not very efficient at producing lensing events.  This
notwithstanding, the optical depth increases by a factor of 5. Panel~(c)
takes the major axis direction maximizing the lensing efficiency, as
determined from the ratio.  Note this occurs when the major axis passes
near the midpoint of the LOS, where $x(1-x)$ is maximized.
However, the LOS passes through lower density contours than in case~(b).
With 35\% less intervening mass, the higher efficiency does not lead to
a significant increase in the optical depth. Panel~(d) takes the major
axis direction maximizing the optical depth.  This direction is the
optimal tradeoff between maximizing mass and maximizing efficiency, so
falls between cases~(b) and~(c).

For these two reasons, more mass and better placement, the triaxial
versions of nearly all the models examined in this paper lead to
larger optical depths than the spherical versions, by roughly
$\sim 20-50\%$ (as shown in the next section, see \reftab{ODResults}).

\subsection{\label{sec:PositionDisc} Position}

An additional consideration taken into account in these models is the
center of the spheroid.  Models H2 \& P2 have the spheroid location
determined from the data, rather than fixed to the Galactic Center.
The center that best fits the data is actually several kpc from the
Galactic Center, approximately 3 kpc along the $y$ axis in
galactocentric coordinates.  This direction is away from the LMC line
of sight, so these shifted spheroid models lead to lower optical
depths.

An off-centered spheroid is not unreasonable from a dynamical
standpoint.  Some excited states of the galaxy involve weakly damped
modes that allow a shifted center to remain for extended periods of
time (M.~Weinberg 2005, private communication).  
Disturbances, including interactions with
nearby galaxies (such as the LMC), can pump this excited state back up.
However, the size of this offset, $\approx 3$ kpc, is larger than the
offsets typically discussed in the field.

The uncertainty in the position of the spheroid center is dominated by
systematics.  Regions of known clumps of stars and contamination are
removed when fitting the SDSS data; the remaining regions are assumed to
represent a smooth spheroid distribution.  However, an unknown
clump/stream of stars (perhaps from another, as yet undetected, galaxy
being absorbed by the Milky Way) could skew the results and lead to a
shift in the fits.  Fits would be particularly sensitive to such
structure at the South Galactic Cap, due to the limited SDSS data in
this region (three stripes).  However, the structure would have to
be significant in size to produce such a shift in the fit-- on the scale
of the tidal streams of the Sagittarius dwarf galaxy
\citep{Majewski:2003ux,Newberg:2003cu} (note all three stripes in the
South Galactic Cap contribute to the shift in the fit).  In addition,
the velocity profiles of stars within these stripes do not give a
noticeable indication of any significant structure in the region.

It has been brought to our attention by 
A.~Gould (2005, private communication) that the
dynamics of a shifted spheroid might lead to a net motion of the local
spheroid population with respect to the galactic rest frame, at odds
with current observations \citep{Gould:2002hf,Gould:2003hc}.  The
shift (as well as a triaxial shape) might also lead to non-trivial cross
terms in the velocity dispersion covariance matrix.
These considerations are not examined in this paper, but will be
investigated in the future.

\subsection{\label{sec:ProfileDisc} Density Profile}

The choice of density profile is also an important factor in the
optical depth prediction.  The profiles that fall off more slowly
with distance from the spheroid center have more mass along the line
of sight and hence a higher optical depth.
For example, the Hernquist fit profile ($r^{-1} (a+r)^{-3}$) falls as
$r^{-1}$ for a significant portion of the LMC line of sight and it does
not reach the more rapid $r^{-4}$ drop until $r \gg a \approx 15$ kpc.
The MACHO profile ($r^{-3.5}$) and power law $F/G$ fit profile
($r^{-\alpha}$, where $\alpha\approx3$), on the other hand, fall rapidly
at all distances.  For a fixed local spheroid density, the Hernquist
profile will yield higher densities at farther distances than the other
profiles and result in a larger optical depth.

\subsection{\label{sec:DensityDisc} Density Normalization}

The factor that has the strongest effect on the optical depth is the
normalization of the spheroid density. We use four normalization
choices.  The lightest local density is due to the GC and MF
normalizations arising from the $F/G$ turnoff stars counts (Newberg/Yanny
data; see \reftab{DerivedParams}), the local density from the MACHO
Collaboration is up to a factor of ten higher, and the OC1 normalization
is by far the largest.

Throughout the paper, we use the local spheroid density and density
profile normalization interchangably.  Indeed, the normalization of the
density profiles (equations~\ref{eqn:MACHOProfile},
\ref{eqn:OC1Profile}~\&~\ref{eqn:HernProfile})
are explicitly defined in terms of the local density $\rho_{S0}$.
Implicitly assumed in these equations, however, is that the spheroid
is composed of a single, smooth component, in which case the $\rho_{S0}$
parameter used to define the normalization in these equations is, in
fact, the actual local density of the spheroid.  If the spheroid is more
complex, such as containing structure or multiple components, the term
``local spheroid density'' becomes ambiguous: the local density used in
the density profile equations represents only the smooth component
described by that profile and does not include additional structure or
spheroid components, while the ``actual'' local density, measured by
local star counts, would include these additional components.  It is
important to note that the $F/G$ models discussed here have their
density profile normalizations directly fixed by the observations;
the local densities discussed for these models are deduced from this
overall normalization and represent only the smooth spheroid component
(the local density is \textit{not} measured).  The MACHO model takes the
observed (actual) local density \refeqn{MACHOnorm} and assumes it to be
entirely representative of the smooth spheroid given by
\refeqn{MACHOProfile}.

In the remainder of this section, we discuss the ways in which the $F/G$
turnoff star normalizations (GC and MF) might be made compatible with
the MACHO Collabaration normalization.

Could the $F/G$ normalization estimates indicate a lighter spheroid
than previously believed?  The high galactic latitude star counts of
\citet{Gould:1998aw} would also suggest this, with a predicted local
density, $6.4 \times 10^{-5} M_{\odot} \, \mathrm{pc}^{-3}$, 2-3 times
smaller than the estimates from local high velocity star observations.
In order for local high velocity stars to give
an inaccurate indication of the spheroid normalization, leading to a
heavier spheroid prediction, the local density must be higher than the
smooth background distribution due to the presence of a ``lump'',
stream, or additional spheroid component.  Indeed, we know that such
streams exist-- the leading edge of the Sagittarius dwarf galaxy
passes very near to the solar neighborhood and could contribute a set
of high proper motion stars~\citep{Majewski:2003ux}.  If this stream
was, however, the dominant component of the local high velocity stars,
it would likely have been noticed in a kinematic analysis of these
stars: their motions would be highly correlated (see, \eg\ 
\citet{Helmi:1999uj}, who find what appears to be a local stream,
but not of a significant density).  Multiple streams might mask this
signature by making the velocities appear more isotropic, leading to a
local density determination that overestimates that of the spheroid.
\citet{Gould:2003hc}, though, argues from a statistical analysis of local
halo stars that the stellar halo is unlikely to be dominated by streams
in our neighborhood.

Another explanation has been proposed by \citet{Sommer-Larsen:1990} and
studied by others \citep{Chiba:2000vu,Siegel:2002vr}: a two component
spheroid containing the traditional approximately spherical distribution
as well as an additional highly flattened component (not to be confused
with the disk, as this component has no net rotation).  This model leads
to approximately equal contributions to the local spheroid density from
the two components; however, the flattened component does not
contribute to non-local star counts such as that of
\citeauthor{Gould:1998aw} and the $F/G$ turnoff stars used in this
paper.  In this case, the local spheroid density estimates in
\citeauthor{Gould:1998aw} and those based on $F/G$
turnoff stars (see \reftab{DerivedParams}) represent only the extended
spheroid contribution and not the flattened component contribution.
Local star counts, which contain both components, would naturally
yield a higher estimate of the local density, by approximately a
factor of 2, and there would be no discrepancy between these two types
of observations.

Alternatively, one can consider the possibility that our mechanisms for
determining the $F/G$ normalizations are yielding underestimates.  The
GC estimate is made by assuming the spheroid stellar population is
similar to that of a globular cluster, yielding similar mass-to-light
ratios.  However, if low mass stars evaporate from globular clusters,
the cluster mass-to-light ratio would be lower than that of the spheroid
and our GC normalization estimate would then likewise be low
\citep{Gould:1998aw}.

The MF $F/G$ normalization estimate was determined by using the spheroid
mass function from \citet{Gould:1998aw}.  The \citeauthor{Gould:1998aw}
MF, given by \refeqn{PDMF}, was determined by examining faint (hence,
distant spheroidal) main sequence stars at high galactic latitudes, so
should probably not have been contaminated by a local bump or a second
spheroid component, as discussed above, and should be reasonably valid
for the $F/G$ models (also non-local star counts).  However, inferring a
MF from such star counts is no simple matter, and some uncertainties
exist.  A study of spheroid stellar populations by
\citet{Gould:2002hf} suggests there may be more low mass stars than
accounted for by the \citeauthor{Gould:1998aw} MF
\footnote{The \citeauthor{Gould:2002hf} paper is based upon
kinematically selected local stars and could, in principle, be measuring
a different population of stars (including, \eg, a second spheroid
component as discussed previously) than the non-local observations of
\citeauthor{Gould:1998aw}.}.  In that case, our
$F/G$ normalization estimate would be low.  Both MFs are in agreement
for the number of higher mass spheroid stars and extrapolation to
turnoff star masses predicts a number of turnoff stars consistent with
that observed by the $F/G$ survey.  Thus, the $F/G$ survey does not
conflict with either of these MFs.  Since the disagreement between these
MFs is for low mass stars and the $F/G$ survey only includes (high mass)
turnoff stars, the $F/G$ models do not give us an indication of which MF
is correct \footnote{We thank the referee for bringing this issue to
our attention.}.

We wish to emphasize that our two estimation methods provide only rough
estimates: the factor of 4 difference between them should be an
indication of this.  While these estimates \textit{might} point to a
lighter spheroid mass than that predicted by local star counts (and used
by MACHO), given the uncertainties in our derivations and the issues
discussed above, our estimates should
not be considered to be incompatible with those of the local star
counts.  The $F/G$ estimate using the MF ($\rho_{S0} \sim 5 \times
10^{-5} M_{\odot} \, \mathrm{pc}^{-3}$) is only 2-3 times smaller than
the MACHO local density ($\rho_{S0} = 1.18 \times 10^{-4} M_{\odot} \,
\mathrm{pc}^{-3}$); it is certainly possible that this simply arises
due to approximations made in the estimation process (yielding
underestimates at this level).  As such, we draw no conclusions as to
whether the MACHO and $F/G$ estimated local densities are
incompatible.

\section{\label{sec:Results} Results}

\begin{table*}  
  \caption[Optical Depths]{
    The optical depth due to the Milky Way spheroid (shown in units of
    $10^{-7}$) for the four density profiles described in the text:
    the upper panel shows results for MACHO, OC1, and power law profiles
    while the lower panel shows results for Hernquist profiles.
    Results are presented for both spherical and triaxial spheroids, for
    the two cases for which the spheroid center is located at and
    shifted away from the Galactic Center.
    The spherical results are determined by setting $p\!=\!1$ \&
    $q\!=\!1$, but leaving the remaining parameters as given in
    \reftab{TAParams}.  The local spheroid density $\rho_{S0}$ has been
    determined in three ways (MACHO, GC, and MF) as described in
    \refsec{HernNorm}.
    }
  \label{tab:ODResults}
  \newcolumntype{e}{D{.}{.}{5}}
  \newcolumntype{f}{D{.}{.}{8}}
  \footnotesize
  \hspace*{\stretch{1}}
  \begin{tabular}{llefef}
    \hline\hline
    & & \multicolumn{2}{c}{Traditional Galactic Center}
      & \multicolumn{2}{c}{Shifted Spheroid Center} \\
    & & \multicolumn{1}{c}{Spherical} & \multicolumn{1}{c}{Triaxial}
      & \multicolumn{1}{c}{Spherical} & \multicolumn{1}{c}{Triaxial} \\
    \hline
    \multicolumn{2}{l}{Parameters} \\
    \multicolumn{2}{l}{\qquad $(R_0+\Delta x,\,\Delta y,\,\Delta z)$ (kpc)}
      & \multicolumn{2}{c}{$(8.5,\,0,\,0)$}
      & \multicolumn{2}{c}{$(8.2,\,2.9,\,0.0)$} \\
    \multicolumn{2}{l}{\qquad $(\theta,\,\phi,\,\xi)$}
      & \multicolumn{2}{c}{$(72^\circ,\,-4^\circ,\,14^\circ)$}
      & \multicolumn{2}{c}{$(52^\circ,\,-6.5^\circ,\,16^\circ)$} \\
    \multicolumn{2}{l}{\qquad $(p,\,q)$}
      & \multicolumn{1}{c}{$(1,\,1)$}
      & \multicolumn{1}{c}{$(0.72,\,0.59)$}
      & \multicolumn{1}{c}{$(1,\,1)$}
      & \multicolumn{1}{c}{$(0.74,\,0.66)$} \\
    \hline
    \multicolumn{2}{l}{Optical Depths (in units of $10^{-7}$)} \\[0.75ex]
    \quad MACHO     & ($r^{-3.5}$)
      & 0.030  & 0.044   & 0.021  & 0.025  \\[0.75ex]
    \quad OC1       & ($r^{-3}$)
      & 0.40   & 0.51    & 0.27   & 0.46    \\[0.75ex]
    \quad Power Law & ($r^{-\alpha}$)       
      & & \multicolumn{1}{c}{\textit{(Model P1)}}
      & & \multicolumn{1}{c}{\textit{(Model P2)}} \\
    \multicolumn{2}{l}{\qquad $\rho_{S0}$ (MACHO)}
      & 0.037  & 0.052   & 0.028  & 0.032   \\
    \multicolumn{2}{l}{\qquad $\rho_{S0}$ (GC)}  
      &        & 0.0059  &        & 0.0040  \\
    \multicolumn{2}{l}{\qquad $\rho_{S0}$ (MF)}  
      &        & 0.024   &        & 0.016   \\
    \hline\hline
    \\[-1.0ex]
    \hline\hline
    \multicolumn{2}{l}{Parameters} \\
    \multicolumn{2}{l}{\qquad $(R_0+\Delta x,\,\Delta y,\,\Delta z)$ (kpc)}
      & \multicolumn{2}{c}{$(8.5,\,0,\,0)$}
      & \multicolumn{2}{c}{$(8.2,\,2.9,\,0.0)$} \\
    \multicolumn{2}{l}{\qquad $(\theta,\,\phi,\,\xi)$}
      & \multicolumn{2}{c}{$(70^\circ,\,-4.5^\circ,\,14^\circ)$}
      & \multicolumn{2}{c}{$(48^\circ,\,-8^\circ,\,12^\circ)$} \\
    \multicolumn{2}{l}{\qquad $(p,\,q)$}
      & \multicolumn{1}{c}{$(1,\,1)$}
      & \multicolumn{1}{c}{$(0.73,\,0.60)$}
      & \multicolumn{1}{c}{$(1,\,1)$}
      & \multicolumn{1}{c}{$(0.73,\,0.67)$} \\
    \hline
    \multicolumn{2}{l}{Optical Depths (in units of $10^{-7}$)} \\
    \quad Hernquist & [$r^{-1} (a+r)^{-3}$] 
      & & \multicolumn{1}{c}{\textit{(Model H1)}}
      & & \multicolumn{1}{c}{\textit{(Model H2)}} \\
    \multicolumn{2}{l}{\qquad $\rho_{S0}$ (MACHO)}
      & 0.047  & 0.059   & 0.036  & 0.035   \\
    \multicolumn{2}{l}{\qquad $\rho_{S0}$ (GC)}  
      &        & 0.0055  &        & 0.0033  \\
    \multicolumn{2}{l}{\qquad $\rho_{S0}$ (MF)}  
      &        & 0.022   &        & 0.013   \\
    \hline
  \end{tabular}
  \normalsize
\end{table*}

We have examined four spheroid models:
\begin{enumerate}
  \setlength{\itemsep}{0pt}
  \item a profile $\rho(r) \propto r^{-3.5}$ used by the MACHO
        collaboration and based upon local high velocity star counts
        and high galactic latitude star counts,
        see \refeqn{MACHOProfile};
  \item a profile $\rho(r) \propto r^{-3}$ referred to as OC1 and
        obtained via galactic dynamics, see \refeqn{OC1Profile};
  \item a new model: a true Hernquist profile of \refeqn{HernProfile},
        with $\alpha=1$ \& $\delta=3$, obtained by fits to spheroid
        $F/G$ turnoff stars; and
  \item a power law profile of \refeqn{HernProfile}, $\delta=0$, also
        obtained by fits to spheroid $F/G$ turnoff stars.
\end{enumerate}
For all four of the models, we have allowed for the possibility of
triaxiality by replacing $r$ with $r_{\TA}$ in the profiles, defined in
\refeqn{triaxialr}.  We have also allowed the spheroid center
to move away from the Galactic Center.  Our results are shown in
\reftab{ODResults}.  We will now discuss each of the models in turn.

\subsection{MACHO Model}

{\it Normalization:} Since the density profile \refeqn{MACHOProfile} in
the MACHO model is based partly on local star counts, which presumably
gives a reliable indication of the local spheroid density regardless of
the spheroid model, we use the local spheroid density given in
\refeqn{MACHOnorm} as the local density in the triaxial case as well.

{\it Triaxiality and position:} In \reftab{TAParams}, we obtained the
best fits to the Newberg/Yanny data for the triaxility and central
spheroid position for arbitrary power law density profiles (models P1
and P2). Though the best fit power law index in the Table ($\rho \propto
r^{-\alpha}$) was closer to $\alpha=3.0$ rather than the $\alpha=3.5$
of the MACHO profile, still we feel confident that we can use the
results of models P1 and P2 in \reftab{TAParams} as a good estimate to
triaxiality and position for the MACHO case as well.

{\it Results for Spheroid Contribution to Optical Depth:} As shown in
\reftab{ODResults}, triaxiality boosts the spheroid optical depth of
this profile by $(25-50)\%$ for either choice of the spheroid center.
However, shifting the spheroid center suppresses the optical depth as
the densest portions of the spheroid are moved away from the LMC line
of sight.  In addition, the highest densities along the line of sight
are closer to the local position, where the microlensing cross-section
is smaller [Note this corresponds to $x \approx 0$ in \refeqn{od}, so
  the integrand is small even where the density is large].  The
combination of these two modifications (spheroid shape and position)
results in a decrease of the spheroid contribution to the optical
depth from $0.030 \times 10^{-7}$ in the traditional model to $0.025
\times 10^{-7}$, a drop of $\sim 20\%$. Since the spheroid
contribution to the overall optical depth was only a few percent in
the spherical MACHO case, this change is unimportant to the estimate
of the number of MACHOs in the Halo of our Galaxy.

\subsection{Dynamical Models (OC1)}

{\it Normalization:} Dynamical models are based upon rotation curves
and are sensitive to the spheroid mass.  Thus for the OC1 profile
(\refEqn{OC1Profile}), we fix the spheroid mass (rather than the local
density) when modifying the model for a triaxial shape.

{\it Triaxility and Position:} Again, the spheroid position and
triaxiality are taken from models P1 \& P2 in \reftab{TAParams}.  

{\it Results for Spheroid Contribution to Optical Depth:} 
Triaxiality gives a $(25-70)\%$ boost in the optical depth, but again
the shifting of the spheroid suppresses the optical depth, resulting in
an overall increase from $0.40 \times 10^{-7}$ to $0.46 \times 10^{-7}$
when both effects are taken into account. The much heavier spheroid
predicted by this model results in a sizable contribution to the optical
depth in all cases.  For the original spherical OC1 spheroid, 50\% of
the excess optical depth to the LMC was due to the spheroid, so that
the maximum MACHO contribution to the dark matter of the Milky Way
Halo was roughly 10\%.  With the triaxiality taken into account,
the new number would be roughly 8\%.  However, since the errors on 
the  observed optical depth from the MACHO experiment to 1$\sigma$ are
as high as 30\%, unfortunately the variation in the optical depth
between the spherical and triaxial cases is smaller that the
uncertainties in the observed optical depth.

\subsection{New spheroid model:  Hernquist profile with Newberg/Yanny
            fits}

Our third set of models uses the Hernquist profile with fits to the
Newberg/Yanny data.  Model H1 has a triaxial spheroid with spheroid
centered at the traditional Galactic Center.  Model H2 has a triaxial
spheroid with a shifted center (not at the Galactic Center).  For
both H1 and H2 we take the values of core radius obtained in
\reftab{TAParams} from fits to the Newberg/Yanny data.

{\it Normalization:} We use three different normalizations for the
density profiles in \refeqn{HernProfile}.  First, we assume the MACHO
value for the local density given in \refeqn{MACHOnorm}. Second,
we use the normalization from the globular cluster comparison (GC).
Third, we use the normalization obtained using the spheroid mass
function (MF). These latter two cases, discussed previously in Section
\ref{sec:HernNorm}, do not apply to the spherical models; hence no
results for spherical models with GC or MF normalization are
presented.

{\it Triaxiality and Position:} The triaxiality and position of the
spheroid center for models H1 and H2 can be found in \reftab{TAParams}.

{\it Results:} With the local density fixed to the MACHO value (MACHO
norm\-al\-iz\-ation), the Hernquist profile optical depth increases with
triaxiality when the spheroid is fixed to the Galactic Center (model
H1).  The contribution to the optical depth is then still only $\sim
5\%$ of the observed value, leading to at most one or two
of the observed microlensing events.
However, when the spheroid is shifted as well (model H2), the
overall optical depth decreases.  This effect is partly due to the
fact that, in model H2, our local position is so near the major axis
that no dense region ``appears'' in the LMC line of sight when going
from a spherical to triaxial shape.  In fact, our local position is
close to the densest portion of the spheroid, so that with a fixed
local density the overall mass of the spheroid decreases.  If we use
the GC or MF normalizations for the spheroid density profile, the
optical depth becomes so small as to be irrelevant.

\subsection{Power Law Model fit to Newberg/Yanny data}

Our fourth set of models uses a power law profile with fits to the
Newberg/Yanny data.   Model P1 has a triaxial spheroid with spheroid
centered at the traditional Galactic Center.  Model P2 has a triaxial
spheroid with a shifted center (not at the Galactic Center).
We use the same three normalizations for the density profiles as
in the third set of models: MACHO, GC, and MF.  The triaxiality
and position of the spheroid center for models P1 and P2 can be found
in \reftab{TAParams}.
 
{\it Results:} Our results for the fourth set of models depend on the
choice of normalization of the density profile.  Using the MACHO
normalization for models P1 and P2 with triaxiality, we obtain very
similar results to those found with the MACHO profile.
This similarity is expected since the MACHO profile is also
a power law.  The differences stem from the best fit power $\rho
\propto r^{-\alpha}$, where $\alpha = 3.0$ for P1 and $2.9$ for P2, as
opposed to $\alpha = - 3.5$ assumed by the MACHO group.  Because the
power law drops a bit more slowly, the fits yield $\sim 25\%$ larger
optical depths.  The shifted spheroid center again suppresses the
optical depth, yielding a net decrease of $\sim 10\%$.

Results are also presented in \reftab{ODResults} for the two other
normalizations of the density profile, GC and MF.  As mentioned above,
these do not apply to the spherical case and hence results are only
presented for the triaxial case.  We find that the optical depth is
$\sim 30\%$ lower when the spheroid center is shifted than when it is
lined up with the Galactic Center.  For both GC and MF normalizations,
the optical depth is too small to be of any significance, $<2\%$ of
the excess MACHO optical depth.

\section{\label{sec:Discussion} Discussion}

Our results of the previous section depend on the triaxiality of the 
spheroid, the location of its center, the density profile, and the 
normalization.  While the importance of these elements has been
emphasized previously, here we demonstrate explicitly the dependence
of our results on these various factors.

\subsection{\label{sec:discusstriax} Triaxiality}

Due to more mass and better placement, the triaxial
versions of nearly all the models examined in this paper lead to
larger optical depths than the spherical versions, with increases of
$\sim 20-50\%$.
The exception is model H2 when the local density is fixed.  In this
model, the major axis is closer to us than any of the other models;
this can be seen by noting that the local triaxial radius,
$R_{0,\TA} = 9.5$ kpc (see \refeqn{triaxialr}), is very close to the
physical distance, $8.8$ kpc.  Then triaxiality boosts the spheroid
density only near us, but points farther away along the line of sight
will fall on lower density contours than they would in spherical
models; compare Figs.~\ref{fig:MADirection}(a)
\&~\ref{fig:MADirection}(b) in our previous example.  Triaxiality
does not as significantly increase the optical depth in this case (the
fixed normalization also plays a part, as mentioned previously).

Clearly, the direction of the major axis matters.  In fact, the
observed direction in most of the models is rather fortunate as far as
optimizing the optical depth.  The optical depth is maximized for a
major axis passing through the line of sight.  While the observed
direction does not exactly do this, it passes closer to the line of
sight than most other directions.  In addition, the optical depth is
maximized for a major axis near to us (but not too near); the observed
direction is reasonably optimal in this regard.

\subsection{\label{sec:discussposition} Position}

The location of the spheroid center also affects our results.
Models H2 \& P2 have the spheroid location
determined from the data, rather than fixed to the Galactic Center.
The center that best fits the data is actually several kpc from the
Galactic Center, approximately 3 kpc along the $y$ axis in
galactocentric coordinates.  This direction is away from the LMC line
of sight, so these shifted spheroid models lead to lower optical
depths, by $\sim 25-40\%$ for a fixed local density, as shown in the
previous section.  This effect generally goes in the opposite direction to
that of triaxiality; the combination of a shifted center and triaxial
shape leads only a modest increase ($\!\!\lae 10\%$) or decrease
($\!\!\lae 25\%$) in the optical depth for the density profiles
considered.

\subsection{\label{sec:discussprofile} Density Profile}

Our results for the optical depth also depend on the choice of density
profile.  To illustrate the impact of the density
profile, let us compare the MACHO profile ($r^{-3.5}$), power law
$F/G$ fit profile ($r^{-\alpha}$), and Hernquist fit profile ($r^{-1}
(a+r)^{-3}$) for the case where the local density is fixed to the
value used by the MACHO Collaboration.  The MACHO profile yields the
smallest optical depth in this case, while the Hernquist profile
yields the largest.  The reason is the more rapid $r^{-3.5}$ drop in
density with the MACHO profile; for a significant portion of the LMC
line of sight, the Hernquist profile falls as $r^{-1}$ and it does not
reach the more rapid $r^{-4}$ drop until $r \gg a \approx 15$ kpc.
The Hernquist profile, then, results in higher densities at $r > R_0$
(distance from us to Galactic Center) and a higher optical depth.  In
\refsec{Results} we saw that, combined with triaxiality, the
prediction for the spheroid optical depth nearly doubles from the
original MACHO profile $0.030 \times 10^{-7}$ to the Hernquist profile
$0.059 \times 10^{-7}$ (including shifting of the spheroid center
results in only a modest $16\%$ net increase to $0.035 \times
10^{-7}$).  The Hernquist profile provides a better fit to the data
than a power law, with a reduced $\chi^2$ of 1.42 \& 1.37 for models
H1 \& H2, respectively, versus 1.51 \& 1.49 for models P1 \& P2,
respectively.

\subsection{\label{sec:dicsussnorm} Normalization}

By far the largest concern in determining if the spheroid can account
for the MACHO observed optical depth is the normalization of the
models, corresponding to the spheroid mass or local density.  The
MACHO Collaboration used a value of $1.18 \times 10^{-4} M_{\odot} \,
\mathrm{pc}^{-3}$ for the local density, determined from local (high
velocity) and non-local (high galactic latitude) star counts.  Since
the local measurements presumably fix the local density regardless of
the spheroid model, we have examined the optical depth of our power
law (P1 \& P2) and Hernquist (H1 \& H2) profiles using this local
density to fix the normalization.  In these cases, the optical depth
from the spheroid comprises only a relatively small portion of the
excess observed value.  While model H1 demonstrates that an alternative
profile with a triaxial shape can give a relatively large increase
over the traditional spherical power law profiles, the contribution to
the optical depth of $0.059 \times 10^{-7}$ in this case is still only
$\sim 7\%$ of the excess observed value, contributing at most one or two
of the observed events.

For the fits to the $F/G$ star surveys (normalizations GC and MF), we
also estimated the spheroid density profile normalizations directly
from the fit normalizations.  The two different estimation techniques
both predict a local density up to an order of magnitude
smaller than that used by the MACHO Collaboration.  For these
estimates, the spheroid optical depth is truly insignificant, $\lae 3\%$
of the excess observed value.  However, we would like to stress that
these were only quick, non-rigorous estimates and should not be
regarded as highly accurate.

The only model, then, with optical depths of a significant size is
OC1, due to its much heavier spheroid.  Even the original model
predicts $0.40 \times 10^{-7}$ for the optical depth, $\sim 50\%$ of
the excess observed value.  Triaxiality and spheroid position vary
this number by as much as $30\%$ (note the spheroid mass, not the
local density, is fixed for this model).  OC1 with a triaxial shape
would account for $\sim 65\%$ of the excess observed value.
Triaxiality can lead to a significant increase in the optical depth,
but only if the spheroid is heavy and the optical depth is already
significant.

We would like to remind the reader that OC1 and other dynamical models
with heavy spheroids are based upon rotation curves and other
measurements of the inner galaxy.  As such, they may provide accurate
models of the inner portion of the spheroid, but are not necessarily
accurate at the larger radii ($r \gae R_0$) relevant to the LMC
optical depth (note the LMC line of sight does not pass much closer than
$R_0$ to the Galactic Center).  The (local and non-local) star counts
forming the basis of the MACHO local density are taken at $r \sim 5-20$
kpc and are likely more reliable for describing the relevant regions of
the galaxy.  In addition, while the $F/G$ normalization estimates may
not be highly accurate, they would need to be off by almost two orders
of magnitude to be compatible with the OC1 model-- it is not obvious
how these estimates could be off to such a degree.

\section{\label{sec:Summary} Summary}

We have investigated the effects of triaxiality of the spheroid, as
discovered by Newberg and Yanny, on the optical depth to the LMC,
which is used to ascertain the number of MACHOs in our Halo.  We
showed that there are two competing effects.  First, the triaxiality
of the fit allows for a larger spheroid mass along the line of sight
between us and the LMC; the placement of the mass is also optimal for
producing microlensing. This effect leads to a larger optical depth
due to the Milky Way spheroid and would imply the existence of fewer
MACHOs to fit the data.  However, allowing the center of the spheroid
to move (rather than fixing it to coincide with the Galactic Center)
leads to an effect that goes in the other direction: the best fit to
the spheroid center is \textit{away} from the line of sight to the LMC.
As a consequence, these two effects tend to cancel so that the change
in optical depth due to the Newberg/Yanny triaxial halo is smaller
than the errors in the measurement.

We considered four spheroid models corresponding to four different
density profiles. Within these models, we allowed for four different
sets of normalization, with different corresponding local densities,
and obtained the resulting optical depth.

Prior to this work there were two widely diverging results for the
spheroid optical depth: those from the MACHO experiment and the OC1
model discussed in the text.  The reason for the discrepancy is
primarily due to a difference in normalization for the local density.
Compared to the standard spheroid optical depth obtained by the MACHO
experiment, if we continue to use the MACHO normalization and local
density, we find that the increase due to triaxiality is at most a
factor of two: the spheroid optical depth may increase from 1/40 of
the total optical depth toward the LMC to 1/20 of it.  Hence one or
two of the observed events may be due to spheroid microlensing.  If
one considers the alternate heavy OC1 model, triaxiality changes the
spheroid contribution from 1/2 the excess optical depth to 2/3 (within
$\sim 1\sigma$ of the experimental excess).  A third normalization for
the local density derived from $F/G$ starcounts in the Newberg/Yanny
data appears to be more consistent with a lighter spheroid and
predicts only negligible contributions to the optical depth.  For the
four spheroid models we have considered, the MACHO contribution to the
dark matter in the Galactic Halo is then in the (8-20)\% range if all
excess lensing events are assumed to be due to MACHOs, where the maximum
is due to the original MACHO collaboration results and the minimum is
consistent with 0\% at the 1$\sigma$ error level in the data.


\ack
  C.S.\ thanks W.\ Sutherland, M.\ Weinberg and A.~Font for helpful
  conversations.  C.S.\ and H.J.N.\ thank A.~Gould for comments on the
  initial draft of this paper.
  C.S.\ and K.F.\ acknowledge the support of the DOE and the Michigan
  Center for Theoretical Physics via the University of Michigan.
  P.G.'s work was partially supported by NSF grant PHY-0456825.
  H.J.N.\ acknowledges funding from NSF grant AST 03-07571.








\end{document}